\documentclass[runningheads,a4paper]{llncs}

\usepackage{amsmath,graphicx}
\usepackage{amssymb}
\usepackage{mathabx}
\usepackage{comment}
\usepackage{sidecap}
\usepackage[disable]{todonotes} 
\usepackage{multirow}
\usepackage{url}
\usepackage{float}

\input{alphabet}
\input{abrege}
\def\acc#1{\left\{#1\right\}}

\def\bigacc#1{\bigl\{#1\bigr\}}            
\def\bigcro#1{\bigl[#1\bigr]}

\def\diag{{\mathrm{diag}}}

\def\arrayp{\renewcommand{\arraystretch}{.7}\setlength{\arraycolsep}{2pt}}
\def\tabp{\renewcommand{\arraystretch}{.7}\setlength{\tabcolsep}{2pt}}

\newsavebox{\fminibox}
\newlength{\fminilength}
\newenvironment{fminipage}[1][\linewidth]
  {\setlength{\fminilength}{#1}
   \begin{lrbox}{\fminibox}\begin{minipage}{\fminilength}}
  {\end{minipage}\end{lrbox}\noindent\fbox{\usebox{\fminibox}}}

\def\T{^\tD} 
\def\+{^\dagger}
\def\I{\,|\,}           

\def\nequiv{\not\kern-.05em\equiv}
\def\egal{\kern-.5em=\kern-.5em}        
\def\propt{\kern-.2em\propto\kern-.2em} 
\def\pourtt{\forall\,}                  
\def\wh#1{\widehat{#1}}                 

\def\intdouble{\int\kern-0.3em\int}
\def\inttriple{\int\kern-0.3em\int\kern-0.3em\int}

\def\rond#1{\overset{\kern-0.33em~_\circ}{#1}}
\def\rondit[#1]#2{\overset{\kern#1~_\circ}{#2}}

\def\babs{\begin{abstract}}             \def\eabs{\end{abstract}}
\def\barr{\begin{array}}                \def\earr{\end{array}}
\def\bcc{\begin{center}}                \def\ecc{\end{center}}
\def\cl#1{\centerline{#1}}
\def\bdes{\begin{description}}          \def\edes{\end{description}}
\def\bdoc{\begin{document}}             \def\edoc{\end{document}}
\def\ben{\begin{enumerate}}             \def\een{\end{enumerate}}
\def\beqn{\begin{eqnarray}}             \def\eeqn{\end{eqnarray}}
\def\beqnl#1{\beqn\label{#1}}           \def\eeqnl#1{\label{#1}\eeqn}
\def\beqnx{\begin{eqnarray*}}           \def\eeqnx{\end{eqnarray*}}
\def\bseqn{\begin{subeqnarray}}         \def\eseqn{\end{subeqnarray}}
\def\beq#1\eeq{\begin{equation}#1\end{equation}}
\def\bal#1\eal{\begin{align}#1\end{align}}
\def\balx#1\ealx{\begin{align*}#1\end{align*}}
\def\beqx{$$}                           \def\eeqx{$$}
\def\bfig{\protect\begin{figure}}       \def\efig{\protect\end{figure}}
\def\bfigx{\protect\begin{figure*}}     \def\efigx{\protect\end{figure*}}
\def\bfigt{\protect\begin{figurette}}   \def\efigt{\protect\end{figurette}}
\def\bfl{\begin{flushleft}}             \def\efl{\end{flushleft}}
\def\bfr{\begin{flushright}}            \def\efr{\end{flushright}}
\def\bit{\begin{itemize}}               \def\eit{\end{itemize}}
\def\bmi{\begin{minipage}}              \def\emi{\end{minipage}}
\def\bfmi{\begin{fminipage}}            \def\efmi{\end{fminipage}}
\def\bpic{\begin{picture}}              \def\epic{\end{picture}}
\def\bqu{\begin{quote}}                 \def\equ{\end{quote}}
\def\bqun{\begin{quotation}}            \def\equn{\end{quotation}}
\def\bsl{\begin{slide}}                 \def\esl{\end{slide}}
\def\btabb{\begin{tabbing}}             \def\etabb{\end{tabbing}}
\def\btabl{\begin{table}}               \def\etabl{\end{table}}
\def\btablx{\begin{table*}}             \def\etablx{\end{table*}}
\def\btab{\begin{tabular}} 
\def\btabu{\begin{tabular}}             \def\etabu{\end{tabular}}
\def\btabx{\begin{tabular*}}            \def\etabx{\end{tabular*}}
\def\bbib{\begin{thebibliography}{999}} \def\ebib{\end{thebibliography}}
\def\bver{\begin{verbatim}}             \def\ever{\end{verbatim}}
\def\bca{\begin{cases}}                          \def\eca{\end{cases}}

\def\hspd{\hspace*{1cm}}

\def\fleche{{\Large \raisebox{-.05cm}{$\Rightarrow$}}}
\def\carre{\setlength{\shadowsize}{2pt}\raisebox{-.05cm}{\shadowbox{}}}
\def\flechedbl{{\Large \raisebox{-.05cm}{$\Rightarrow$}}}
\def\tiret{{---~~ }}

\def\B{\mbox{\large$\bullet$}}
\def\C{\mbox{\large$\circ$}}
\def\X{\mbox{\large$\times$}}
\def\x{\raisebox{-3pt}{\Huge$\times$}}
\def\cc#1{\setlength{\tabcolsep}{0pt}\btabu{c}#1\etabu}
\def\fleche{{\Large \raisebox{-.05cm}{$\Rightarrow$}}\XS}
\def\fleched{\XS{\Large \raisebox{-.05cm}{$\Leftrightarrow$}}\XS}
\def\fleches{{\Large \raisebox{-.05cm}{$\rightarrow$}}\XS}
\def\dfleches{{\Large \raisebox{-.05cm}{$\leftrightarrow$}}\XS}
\def\upfleche{{\Large \raisebox{-.05cm}{$\uparrow$}}\XS}
\def\Upfleche{{\Large \raisebox{-.05cm}{$\Uparrow$}}\XS}
\def\dofleche{{\Large \raisebox{-.05cm}{$\downarrow$}}\XS}
\def\Dofleche{{\Large \raisebox{-.05cm}{$\Downarrow$}}\XS}
\def\sefleche{{\Large \raisebox{-.05cm}{$\searrow$}}\XS}
\def\swfleche{{\Large \raisebox{-.05cm}{$\swarrow$}}\XS}
\def\tiret{{---~~ }}
\def\Boulet#1{\hsp{\B~#1}}
\def\Bouletdec#1{\hspd{\B~#1}}
\def\Cdec#1{\hspd{\C~#1}}

\def\cl#1{\centerline{#1}}
\def\eqcl#1{\cl{$\displaystyle#1$}}
\def\eq#1{{$\displaystyle#1$}}
\def\cites#1{{\small\cite{#1}}}
\def\arrayp{\renewcommand{\arraystretch}{.7}\setlength{\arraycolsep}{2pt}}
\def\tabp{\renewcommand{\arraystretch}{.7}\setlength{\tabcolsep}{2pt}}

\def\sk{\vskip.1in}
\def\hsp{\hspace*{.5cm}}
\def\hspp{\hspace*{1cm}}
\def\vsps{\vspace*{.6cm}}
\def\vspc{\vspace*{.5cm}}
\def\vsp{\vspace*{.5cm}}
\def\hspn{\hspace*{-.5cm}}
\def\hspnu{\hspace*{-.4cm}}
\def\vspq{\vspace*{.4cm}}
\def\vspd{\vspace*{.2cm}}
\def\vspt{\vspace*{.3cm}}
\def\vspu{\vspace*{.1cm}}
\def\vspn{\vspace*{-.1cm}}
\def\vspnd{\vspace*{-.2cm}}
\def\vspnq{\vspace*{-.45cm}}

\def\eqcl#1{\cl{$\displaystyle#1$}}
\def\eq#1{{$\displaystyle#1$}}
\def\sump{\mathop{\raisebox{-.3ex}{\text{\Large$\Sigma$}}}}
\def\sumpp{\mathop{\raisebox{-.2ex}{\text{\large$\Sigma$}}}}
\def\prodp{\mathop{\raisebox{-.3ex}{\text{\Large$\Pi$}}}}
\def\prodpp{\mathop{\raisebox{-.2ex}{\text{\large$\Pi$}}}}

\def\P{\mbox{\raisebox{5pt}{\scriptsize($P$)}}}
\def\diagou{\mbox{\Large$\diagup$}}
\def\diagod{\mbox{\Large$\diagdown$}}

\def\cites#1{{\small\cite{#1}}}
\def\citesa#1{{\small\cite[\kern-.3em a]{#1}}}

\RequirePackage{color}

\newcommand{\titresl}[2][]{\cl{
\psframebox*[shadow=true,shadowsize=4pt,linestyle=none,fillcolor=lightgray]{~~\uppercase{#1{\footnotesize #2}}~~}}}

\definecolor{midgray}{rgb}{.5,.5,.5}

\def\carregris{\setlength{\shadowsize}{2pt}\setlength{\fboxsep}{0pt}\setlength{\fboxrule}{0pt}\raisebox{-.05cm}{\shadowbox{{\kern-.5pt \color{midgray}$\blacksquare$\kern-1pt}}}}
\def\Carre#1{\hsp{\carregris}~~\textbf{#1}}
\def\Annee#1#2#3{{\color{#2} #1}#3}

\def\carrerouge{\setlength{\shadowsize}{2pt}\setlength{\fboxsep}{0pt}\setlength{\fboxrule}{0pt}\raisebox{-.05cm}{\shadowbox{{\kern-.5pt \color{RougePhil}$\blacksquare$\kern-1pt}}}}

\def\carrebleu{\setlength{\shadowsize}{2pt}\setlength{\fboxsep}{0pt}\setlength{\fboxrule}{0pt}\raisebox{-.05cm}{\shadowbox{{\kern-.5pt \color{BleuPhil}$\blacksquare$\kern-1pt}}}}

\def\Carreb#1{\hsp{\carrebleu}~~\textbf{#1}}
\def\Carrer#1{\hsp{\carrerouge}~~\textbf{#1}}

\def\Carred#1#2{\hspace*{#2cm}{\carregris~~\textbf{#1}}}

\def\incircgris#1{\raisebox{.25mm}{\pscirclebox[framesep=1pt,fillstyle=solid,fillcolor=lightgray,linestyle=none]{\textbf{\white\footnotesize#1}}}}
              \def\incircgrisfonc#1{\raisebox{.25mm}{\pscirclebox[framesep=1pt,fillstyle=solid,fillcolor=darkgray,linestyle=none]{\textbf{\white\scriptsize#1}}}}

 \def\Cgrdec#1{\hspd{{\color{midgray}\B}~#1}}

\definecolor{BleuPhil}{rgb}{0.25,0.25,1}  		
\definecolor{RougePhil}{rgb}{1,0.15,0.15} 
\definecolor{VertPhil}{rgb}{0.25,1,0.25}  
\definecolor{MagentaPhil}{rgb}{0.5,0,0.7}  
\definecolor{NoirPhil}{rgb}{0.,0.,0.}  

\def\Oh{\Omega \hb}

\def\PM{\kern0pt^{\textrm{{\scriptsize PM}}}\kern0pt}
\def\idxk{^{(k)}}
\newcommand{\figc}[2][]
   {\setlength{\tabcolsep}{0pt}\btabu{c}\includegraphics[#1]{#2}\etabu}
\def\yaxis#1{\cc{\rotatebox{90}{{\small #1}}}}
\def\ylabel#1{\setlength{\tabcolsep}{0pt}\btabu{c}\rotatebox{90}{#1}\etabu}

\graphicspath{{./Figures/},{./Figures_real_data/}}

\newcommand{\keywords}[1]{\par\addvspace\baselineskip
\noindent\keywordname\enspace\ignorespaces#1}

\newcommand{\recentChange}[1]{#1}

\begin{document}

\mainmatter

\title{Physiologically ­informed Bayesian analysis of ASL fMRI data}

\author{
Aina Frau-Pascual$^{1,3}$ \and Thomas Vincent$^{1}$ \and Jennifer Sloboda$^{1}$ \and \\
Philippe Ciuciu$^{2,3}$ \and Florence Forbes$^{1}$
}
\authorrunning{A. Frau-Pascual et al.}

\institute{
$^{(1)}$INRIA, MISTIS, Grenoble University, LJK, Grenoble, France \\
$^{(2)}$CEA/DSV/I$^2$BM NeuroSpin center, B\^at. 145, F-91191 Gif-sur-Yvette, France \\
$^{(3)}$INRIA, Parietal, F-91893 Orsay, France
}

\toctitle{toctitle}
\tocauthor{Authors' Instructions}

\maketitle

\begin{abstract}
Arterial Spin Labelling (ASL) functional Magnetic Resonance Imaging (fMRI) data provides a quantitative measure of blood perfusion, that can be correlated to neuronal activation. In contrast to BOLD measure, it is a direct measure of cerebral blood flow. However, ASL data has a lower SNR and resolution so that the recovery of the perfusion response of interest suffers from the contamination by a stronger hemodynamic component in the ASL signal. In this work we consider a model of both hemodynamic and perfusion components within the ASL signal.
 A physiological link between these two components is analyzed and  used for  a more accurate estimation of the perfusion response function in particular in the usual ASL low SNR conditions.
\end{abstract}

\section{Introduction}
\label{sec:intro}

Arterial Spin Labelling (ASL) \cite{Williams92} provides a direct measure of cerebral blood flow (CBF), overcoming one of the most important limitations of Blood Oxygen Level Dependent (BOLD) signal~\cite{Ogawa92}: BOLD contrast cannot quantify cerebral perfusion.  In contrast to BOLD, ASL is able to provide a measure of baseline CBF
as well as quantitative CBF signal changes in response to stimuli presented to any volunteer in the scanner during an experimental paradigm. Hence, ASL enables the comparison of CBF changes between experiments and subjects (healthy vs patients)
making its application to clinics feasible. In addition, ASL signal localization is closer to neural activity. ASL has already been used in clinics in steady-state for instance for probing CBF discrepancy in pathologies like Alzheimer's disease and stroke, but its use in the functional MRI context has been limited so far.
Despite ASL advantages, its main limitation lies in its low Signal-to-Noise Ratio~(SNR), which, together with its low temporal and spatial resolutions, makes the analysis of such data more challenging.

According to~\cite{Hernandez10,Mumford06}, ASL signal has been typically analyzed with a general linear model (GLM) approach, accounting for a BOLD component mixed with the perfusion component. In such a setting both the hemodynamic response function~(HRF or BRF for  BOLD response function)  and perfusion response function~(PRF) are assumed to be the same and to fit the canonical BRF shape. 
In contrast, an adaptation of the Joint-Detection estimation~(JDE) framework~\cite{Vincent09c} to ASL data has been proposed in~\cite{Vincent13b,Vincent13} to separately estimate BRF and PRF shapes, and implicitly consider the control/label effect which, as stated in~\cite{Mumford06}, increases the sensitivity of the analysis compared to differencing approaches. Although this JDE extension provides a good estimate of the  BRF, the PRF estimation remains much more difficult because of the noisier nature of the perfusion component within the ASL signal. 
In the past decade, phy\-siological models have been described to explain the physiological changes caused by neural
activity. In \cite{Friston00b,Buxton04}, neural coupling, which maps neural activity to ensuing CBF, and the {\it Balloon model}, which relates CBF to BOLD signal, have been introduced. These models describe the process from neural activation to the BOLD measure, and the impact of neural activation on other physiological parameters.

Here, we propose to rely on these physiological models to derive a tractable linear link between perfusion and BOLD components within the ASL signal and to exploit this link as a prior knowledge for the accurate and reliable recovery of the PRF shape in functional ASL data analysis. This way, we refine the separate estimation of the response functions in~\cite{Vincent13b,Vincent13} by taking physiological information into consideration.
The structure of this paper goes as follows: the physiological model and its linearization  to find 
the PRF/BRF link are presented in section~\ref{sec:physio_model}. Starting then from the ASL JDE model  described in 
section~\ref{sec:physio_JDE}, we extend the estimation framework to account for the physiological link in
section~\ref{sec:physio_prior}. Finally, results on artificial and real data 
are presented and discussed in sections~\ref{sec:results_sim}-\ref{sec:discussion}.

\section{A physiologically informed  ASL/BOLD link}
\label{sec:physio_model}

Our goal is to derive an approximate physiologically informed relationship between the perfusion and hemodynamic response functions so as to 
improve their estimation in a JDE framework \cite{Vincent13b,Vincent13}. We show in this section that, although this relationship is an imperfect link resulting from a linearization, it provides a good approximation and allows to capture important features such as a shift in time-to-peak from one response to another. For a physiologically validated model, we use the extended balloon model described below.

\subsection{The extended Balloon model}

The Balloon model was first proposed in \cite{Buxton98} to link neuronal  and vascular processes by considering the capillary as a  balloon that dilates under the effect of blood flow variations. More specifically,  the model describes how, after some
stimulation, the local blood flow $\fb_{in}(t)$ increases and leads to the subsequent augmentation of the local capillary volume $\nub(t)$.  This incoming blood is strongly oxygenated but only part of the oxygen is  consumed.  It follows a local decrease of the deoxyhemoglobin concentration $\xib(t)$ and therefore a BOLD signal variation. 
The Balloon model was then extended in \cite{Friston00b} to include the effect of the neuronal activity $\ub(t)$ on the variation of some auto-regulated flow inducing signal $\psib(t)$ so as  to eventually link neuronal to hemodynamic activity. 
The global physiological model corresponds then to a non-linear system with four state variables $\{ \psib, \fb_{in}, \nub, \xib  \}$ corresponding to normalized flow inducing signal, local blood flow, local capillary volume, and deoxyhemoglobin concentration. Their interactions over time are described by the following system of differential equations:
\bal
\begin{cases}
\frac{d \fb_{in}(t)}{dt} = \psib(t)
\\ 
\frac{d \psib(t)}{dt} =    \eta u(t) 
						- \frac{\psib(t)}{\tau_{\psi}} 
						- \frac{\fb_{in}(t) - 1}{\tau_f}
						\\
			\frac{d \xib(t)}{dt} = \frac{1}{\tau_{m}} \left(\fb_{in}(t) \frac{1 - (1 - E_0)^{1/\fb_{in}(t)}}{E_0} - \xib(t) \nub(t)^{\frac{1}{\tilde{w}}-1} \right)
\\ \frac{d \nub(t)}{dt} = \frac{1}{\tau_{m}} \left(\fb_{in}(t) - \nub(t)^{\frac{1}{\tilde{w}}} \right) \label{equadiff}
\end{cases}
\eal
with initial conditions $\psib(0)=0, \fb_{in}(0)=\nub(0)= \xib(0)=1$. Lower case notation is used for normalized functions by convention. The system depends on 5 hemodynamic parameters:
$\tau_{\psi}$, $\tau_{f}$ and $\tau_m$ are time constants  respectively for signal 
decay/elimination, auto-regulatory feedback from blood flow and mean transit time, $\tilde{w}$  reflects the ability of the vein to eject blood, and $E_0$ is the oxygen  extraction fraction. Another parameter
$\eta$ is the neuronal efficacy weighting term that models neuronal efficacy variability. 

Once the solution of the previous system is found, Buxton et al \cite{Buxton98} proposed the following  expression that links the BOLD response $\hb(t)$ to the physiological quantities considering intra-vascular and extra-vascular components:
\bal
\hb(t) 
& = V_0 [k_1(1-\xib(t)) + k_2(1-\frac{\xib(t)}{\nub(t)}) + k_3(1-\nub(t)) ] \label{hdef}
\eal
where $k_1$, $k_2$ and $k_3$ are
 scanner-dependent constants and $V_0$ is the resting blood volume fraction. According to \cite{Buxton98},${k_1 \cong 7 E_0}$, ${k_2 \cong 2}$ and ${k_3 \cong 2E_0 - 0.2}$ at a field strength of 1.5T and echo time $TE=40$ms. 
 
\begin{figure}[tbm]
	\centering
	\includegraphics[width=0.45\textwidth]{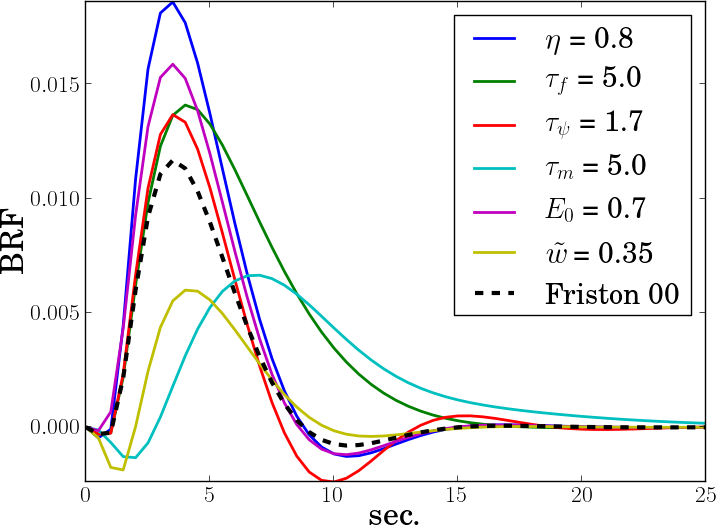}
	\hspace{0.5cm}
	\includegraphics[width=0.45\textwidth]{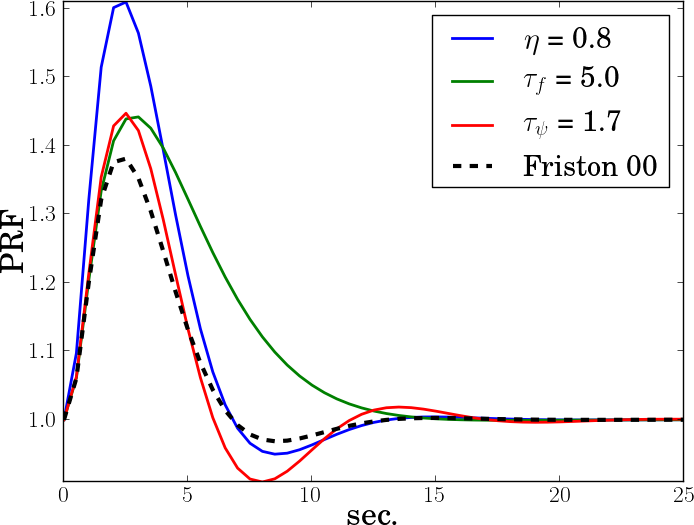}
	\caption{Effect of the physiological parameters on the BRF (left) and PRF (right) shapes. The parameters values proposed in \cite{Friston00b} are used except for one parameter whose identity and value is modified as indicated in the plot.}
	\label{fig:physio_params}
  \end{figure}

The physiological parameters used are the ones proposed by Friston et al in \cite{Friston00b}: 
$V_0=0.02$, $\tau_{\psi}=1.25$, $\tau_{f}=2.5$, $\tau_m=1$,  $\tilde{w}=0.2$, $E_0=0.8$ and $\eta = 0.5$. The BRF and PRF generated using these parameters with the physiological model are shown in Fig.~\ref{fig:physio_params} under the label ``Friston 00" (dashed line).
The rest of the curves show the effect of changing the physiological parameters: $\eta$ is a scaling factor and causes non-linearities above a certain value; $\tau_{\psi}$ controls the signal decay, which is more or less smooth;  the auto-regulatory feedback $\tau_{f}$ regulates the undershoot; the transit time $\tau_m$ expands or contracts the signal in time; the windkessel parameter  $\tilde{w}$  models the initial dip and the response magnitude; the oxygen extraction  $E_0$  impacts the response scale. After analysing the behaviour of the model when varying the parameters values, the impact of each parameter was investigated and we concluded that the values proposed in \cite{Friston00b} seemed reasonable.

\subsection{Physiological linear relationship between response functions}

From the system of equations previously defined, we derive an approximate relationship between the PRF, namely $\gb(t)$, and the BRF, which is given by $\hb(t)$ when $\ub(t)$ is an impulse function. Both BRF and PRF are percent signal changes, and we consider $\gb(t)=\fb_{in}(t)-1$, as $\fb_{in}(t)$ is the normalized perfusion, with initial value $1$. Therefore the state variables are ${\{ \psib, \gb, 1-\nub, 1-\xib \}}$.

In the following we will drop the time index $t$ and consider functions $\hb, \psib, etc.$ in their 
discretized vector form. We can obtain a simple relationship between $\hb$ and $\gb$ by linearizing the system of equations.
 Equation (\ref{hdef}) can first  be linearized into:
\bal
\hb &= V_0[(k_1 + k_2) (1-\xib) + (k_3 - k_2) (1-\nub)] \; .
\eal
We then linearize the system (\ref{equadiff}) around the resting point $\{  \psib, \gb, 1-\nub, 1-\xib \} = \{ \mathbf{0}, \mathbf{0}, \mathbf{0}, \mathbf{0} \}$ as in \cite{Khalidov11}. From this linearization, denoting by 
 $\mathcal{D}$ the first order differential operator and $\Ib$ the identity matrix, we get:
\bal
\begin{cases}
\mathcal{D} \{ \gb \} = - \psib \\ 
\left( \mathcal{D} + \frac{\Ib}{\tilde{w}\tau_{m}} \right) \{ 1-\nub \} = - \frac{1}{\tau_{m}} \gb \\
\left( \mathcal{D} + \frac{\Ib}{\tau_{m}} \right) \{ 1-\xib \} = - \left( \gamma \Ib - \frac{1 - \tilde{w}}{\tilde{w} \tau_m^2} \left( \mathcal{D} + \frac{\Ib}{\tilde{w}\tau_{m}} \right)^{-1} \right) \gb
\end{cases},
\eal
where $\gamma = \frac{1}{\tau_{m}} \left( 1 + \frac{(1-E_0) \ln(1-E_0)}{E_0} \right)$. 
It follows a linear link between $\hb$ and $\gb$ that we write as $\gb = \Omegab \hb$ where:
\bal
\Omegab &=
 V_0^{-1} \left( - (k_1 + k_2) 
  \gamma \Bb + (k_1 + k_2) \frac{1 - \tilde{w}}{\tilde{w} \tau_m^2} \Bb \Ab 
 -  \frac{k_3 - k_2}{\tau_m} \Ab \right)^{-1}  \\
& \text{with } \Ab =  \left( \mathcal{D} + \frac{\Ib}{\tilde{w}\tau_m} \right)^{-1}  \text{ and  }  \ \ \Bb =  \left( \mathcal{D} + \frac{\Ib}{\tau_m} \right)^{-1}
\eal

\begin{figure}[tbm]
	\centering
	\includegraphics[width=0.45\textwidth]{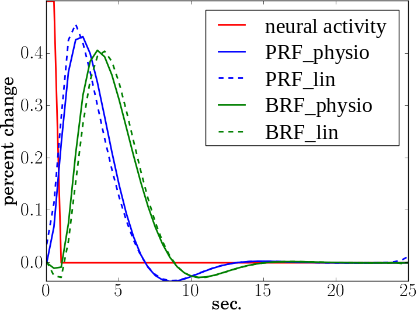}
	\caption{Physiological responses generated with the physiological model, using parameters proposed in \cite{Friston00b}: neural activity $\psib$,  physiological ($\hb_{physio}$ or BRF$_{physio}$) and linearized ($\hb_{lin}$ or BRF$_{lin}$) BRFs, physiological ($\gb_{physio}$ or PRF$_{physio}$) and linearized ($\gb_{lin}$ or PRF$_{lin}$) PRFs.}
	\label{fig:linear_operator}
  \end{figure}

Using values of physiological constants as proposed in~\cite{Friston00b}, 
 Fig.~\ref{fig:linear_operator} shows the BRF and PRF results that we get ($\hb_{lin}$, $\gb_{lin}$) by 
 applying the linear operator to physiologically genera\-ted PRF ($\gb_{physio}$) or 
 BRF ($\hb_{physio}$): ${\hb_{lin} = \Omegab^{-1} \gb_{physio}}$ or 
 ${\gb_{lin} = \Omegab \hb_{physio}}$ compared to these 
 physiologically generated $\hb_{physio}$ and $\gb_{physio}$ 
 functions, computed by using the physiological model differential equations. Note that, although time-to-peak (TTP) values are not exact, the linear 
 operator maintains the shape of the functions and satisfyingly captures the main features of  the two responses. We considered a finer temporal resolution than TR for $\Omegab$ and, besides this, there is no direct dependence on the TR.

The derivation of this linear operator gives us a new tool for analyzing the ASL signal, although this link is subject to caution as linearity assumption is strong and this linearization induces approximation error. 

\section{Bayesian hierarchical model for ASL data analysis}
\label{sec:physio_JDE}

The ASL JDE model described in \cite{Vincent13b,Vincent13}
assumes a partitioned brain into several
functional homogeneous parcels each of which gathers 
signals which share the same response shapes.
In a given parcel $\Pc$, the generative model for ASL time series, measured at times $(t_n)_{n=1{:}N}$ where $t_n = n
\text{TR}$, $N$ is the number of scans and $\text{TR}$ the time of
repetition, with $M$ experimental conditions, reads $\pourtt j \in \Pc$, $|\Pc|=J$:
\bal\label{forward_model}
\yb_j &\!=\!  \sum_{m=1}^M \underbrace{a_j^m\Xb^m \hb}_{(a)} + \underbrace{c_j^m \Wb \Xb^m \gb}_{(b)} + 
\underbrace{\Pb \ellb_j}_{(c)} \!+\! \underbrace{\alpha_j\wb}_{(d)} \!+\! \underbrace{\bb_j}_{(e)} \eal
The signal is decomposed into (a) task-related BOLD and (b) perfusion components given by the first two terms respectively; (c) a drift component $\Pb \ellb_j $ already considered in the BOLD JDE~\cite{Vincent09c}; (d) a perfusion baseline term
$\alpha_j\wb $ which completes the modelling of the perfusion component; and (e) a noise term. 

ASL fMRI data consists in the consecutive and alternated acquisitions of control and magnetically tagged images. The tagged image embodies a perfusion component besides the BOLD one, which is present in the control image too. The BOLD component is noisier compared to standard BOLD fMRI acquisition. The control/tag effect is implicit in the ASL JDE model with the use of matrix $\Wb$.
More specifically, we further describe each signal part below.

\noindent {\bf (a) The BOLD component: } $\hb\in\mathbb{R}^{F+1}$ represents the unknown BRF shape, with size $F+1$ and constant within $\Pc$. 
The magnitude of activation or BOLD response levels  are $\ab=\acc{a_j^m,j\in \Pc, m=1:M}$.

\noindent {\bf (b) The perfusion  component: } It represents the variation of the perfusion from the baseline when there is task-related activity. 
$\gb\in\mathbb{R}^{F+1}$ represents the unknown PRF shape, with size $F+1$ and constant within $\Pc$. 
The magnitude of activation or perfusion 
response levels are $\cb=\acc{c_j^m,j\in \Pc, m=1:M}$. $\Wb$ models the control/tag effect in the perfusion component, and it is further explained below.
 
\noindent {\bf (a-b)} Considering $\Delta t < TR$ 
the sampling period of $\hb$ and $\gb$, whose temporal resolution is assumed to be the same, $\Xb=\{x^{n-f\Delta t}, n=1:N, f=0:F\}$ is a binary matrix that encodes the lagged onset stimuli.
In \cite{Vincent13b,Vincent13}, BRF and PRF shapes follow prior 
Gaussian distributions $\hb \thicksim \mathcal{N}(\zerob,v_\hb \Sigmab_\hb)$ and $\gb \thicksim \mathcal{N}(\zerob, v_\gb \Sigmab_\gb)$,
with covariance matrices $\Sigmab_h$ and $\Sigmab_g$ encoding a constraint on the second order derivative so as to account for temporal smoothness.  
The BOLD (BRLs) and perfusion
(PRLs) response levels (resp. $\ab$ and $\cb$) are assumed to follow different spatial Gaussian
mixture models but governed by
common  binary hidden Markov random fields  $\{q_j^m , j \in \Pc\}$ encoding voxels' activation ($q_j^m=1, 0$ for activated, resp. non-activated)
states for each experimental condition $m$. This way, BRLs and PRLs are independent conditionally to $\qb$: $p(\ab,\cb \I \qb)$. An Ising model on $\qb$ introduces spatial correlation as in~\cite{Vincent13b,Vincent13}. For further interest please refer to \cite{Vincent09c}.
Univariate Gamma/Gaussian mixtures were used instead in \cite{Makni06b} at the expense of computational cost. The introduction of spatial modelling through hidden Markov random fields gave an improved sensitivity/specificity compromise.

\noindent {\bf (c) The drift term:}  It allows to account for a potential drift and any
other nuisance effect ({e.g.} slow motion parameters).
Matrix $\Pb\!=\!\bigcro{\pb_1,\ldots,\pb_{O}}$ 
of size $N \times O$
comprises the values of an orthonormal basis~(\ie $\Pb\T\Pb=\Ib_{O}$). Vector $\ellb_j=(\ell_{o,j}, o=1:O)\T$ 
contains the corresponding unknown regression coefficients for voxel $j$. The prior reads $\ellb_j \thicksim \mathcal{N}(0,v_{\ellb} I_O)$.

\noindent {\bf (b-d) The control/tag vector $\wb$ ($N$-dimensional): }
It encodes the difference in magnetization signs 
between control and tagged ASL volumes.
$w_{t_n}=1/2$ if $t_n$ is even~(control volume) and $w_{t_n}=-1/2$
 otherwise~(tagged volume), and $\Wb=\diag{(\wb)}$ is 
 the diagonal matrix with $\wb$ as diagonal entries. 
 
\noindent {\bf (d) The perfusion baseline:} It is encoded by $\alpha_j$ at voxel $j$. The prior reads $\alpha_j \sim {\cal N}(0, v_{\alpha})$. 

\noindent {\bf (e) The noise term:} It is assumed  white Gaussian with unknown variance $v_{b}$,
$\bb_j\!\sim\!\Nc(\zerob,v_{b}\Ib_N)$.

\noindent{\bf Hyper-parameters $\Thetab$.} 
Non-informative Jeffrey priors are adopted for $\bigacc{v_\bb,
v_{\ellb},$ $v_\alphab}$ and proper conjugate priors
are considered for the mixture parameters of BRLs ($\thetab_\ab$) and PRLs ($\thetab_\cb$). 

\section{A physiologically informed 2-steps inference procedure}
\label{sec:physio_prior}

The BOLD component is known to have a higher SNR than the perfusion component in the ASL signal, and can be estimated with a higher confidence. The link $\gb= \Omegab \hb$  that we derived between both components can then 
be used to inform the PRF from the BRF.  Using this link the other way around may not be satisfying as it may result in a contamination of $\hb$ by a noisier $\gb$. 

This effect has been noticed in the implementation of a physiologically informed Bayesian procedure, considering the generative model (\ref{forward_model}), and the following priors for the PRF and BRF  
$\hb \sim \Nc (0, v_{\hb} \Sigmab_h)$ and $\gb | \hb \sim \Nc (\Omegab \hb , v_{\gb} \Sigmab_g)$, with $\Sigmab_h = \Sigmab_g =  (\Delta t)^4  (\Db_2^t \Db_2)^{-1}$. 
$\Db_2$  is the truncated second order finite difference matrix  of size $(F-1)\times (F-1)$ that introduces temporal smoothness, as in \cite{Vincent13b,Vincent13}, and $v_\hb$ and $v_\gb$ are scalars that we set manually. As seen in Fig.~\ref{fig:jde_simulated}[Middle], this approach does not yield satisfying results, not only for the perfusion component, but also for the BOLD one, compared to the model presented in \cite{Vincent13b,Vincent13}. 

We therefore  propose to exploit the described physiological link in a two-step procedure, in which we first identify hemodynamics properties ($\hat{\hb}$, $\hat{a}_j^m$), and then use the linear operator $\Omegab$ and the previously estimated hemodynamic properties to recover the perfusion component ($\hat{\gb}$, $\hat{c}_j^m$). This way, we avoid an arising contaminating effect of $\gb$ on the estimation of $\hb$, as in the one-step approach in Fig.~\ref{fig:jde_simulated}[Middle]. 
Each step is based on a Gibbs sampling procedure as in \cite{Vincent13b,Vincent13}.\\

\subsection{Hemodynamics estimation step $\mathcal{M}_1$} 

In a first step $\mathcal{M}_1$, our goal is to extract the hemodynamic components and the drift term from the ASL data. In the JDE framework (\ref{forward_model}), it amounts to initially considering the perfusion component as a generalized perfusion term, 
including perfusion baseline and event-related perfusion response.
The generative model (\ref{forward_model}) for ASL time series can be equivalently written, by grouping the perfusion terms involving $\Wb = diag(\wb)$, as 
\bal\label{forward_model2}
\yb_j \!=\!  \sum_{m=1}^M \!a_j^m\Xb^m \hb + \Pb \ellb_j \!+ \Wb \left( \sum_{m=1}^M c_j^m \Xb^m \gb \!+\! \alpha_j\unb \right) \!+\! \bb_j
\eal 
where we consider $\alpha_j \wb = \Wb \alpha_j \unb$. Note that the hemodynamics components BRF $\hb$ and the drift term $\ellb_j$ can be estimated first, by segregating them from a general perfusion term and a noise term. However, the perfusion component is considered in the residuals, so as to properly estimate its different contributions in a second step $\mathcal{M}_2$. 

Given the estimated 
$\wh{\hb}^{\mathcal{M}_1}$, $\wh{\ellb}^{\mathcal{M}_1}$ and $\wh{a}^{\mathcal{M}_1}$, we then compute residuals {$\rb^{\mathcal{M}_1}$} containing the remaining  perfusion component: 
\bal
\rb_j^{\mathcal{M}_1}  = \yb_j - \sum_{m=1}^{M}  \wh{a}_j^{m,\mathcal{M}_1} \Xb^m 
          \wh{\hb}^{\mathcal{M}_1} -\Pb \wh{\ellb}_j^{\mathcal{M}_1}
\eal

\subsection{Perfusion response estimation step ${\mathcal{M}_2}$} 

From the residuals of the first step $\rb^{\mathcal{M}_1}$, we  estimate the perfusion component.
The remaining signal is, according to (\ref{forward_model}), $\forall j = 1:J$,
\beq
\yb_j^{\mathcal{M}_2} = \rb_j^{\mathcal{M}_1} = \sum_{m=1}^{M} c_j^m \Wb \Xb^m \gb + \alpha_j \wb + \bb_j
\eeq
In this step, we introduce a  prior on $\gb$, to account for the already described physiological relationship ${\gb = \Omegab \hb}$: 
\bal
\gb | \wh{\hb}^{\mathcal{M}_1} &\sim \Nc (\Omegab \wh{\hb}^{\mathcal{M}_1}, v_{\gb} \Sigmab_g), \text{ with } \Sigmab_g = \Ib_F \; .
\eal

The significance of the 2-step approach is to first preprocess the data to subtract the hemodynamic component within the ASL signal, as well as the drift effect, and to focus in a second step on the analysis of the smaller perfusion effect.
In \cite{Mumford06}, differencing methods were used to subtract components with no interest in the perfusion analysis and 
directly analyse the perfusion effect in the time series. 
In contrast to these methods, 
we expect to disentangle perfusion from BOLD components by identifying all the components contained in the signal, and to recover them more accurately.

\section{Simulation results}
\label{sec:results_sim}

\begin{figure}[tbm]
  \center
   \figc[height=1.8cm]{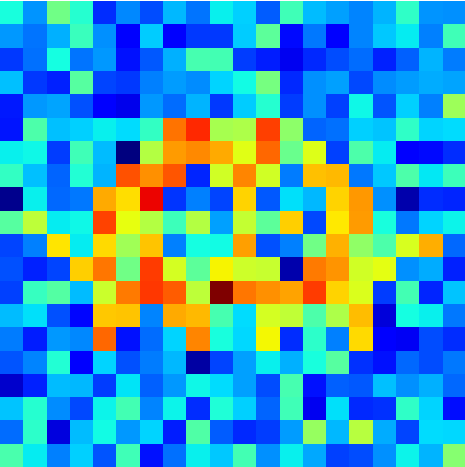}
   \figc[height=1.8cm]{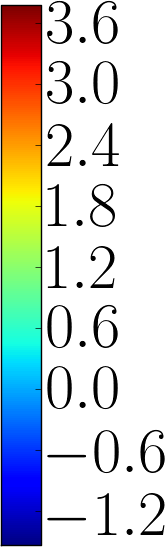}
   \hspace{1cm}
   \figc[height=1.8cm]{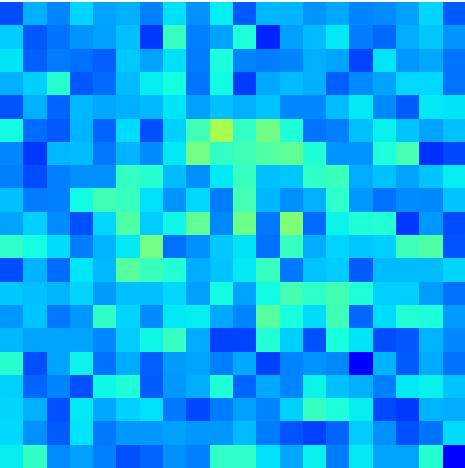}
   \figc[height=1.8cm]{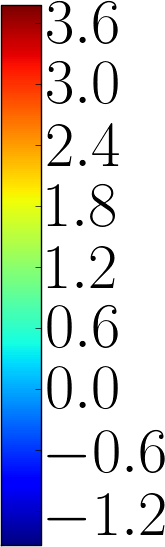}
   \vspace{-0.2cm}
  \caption{BRL and PRL ground truth for a noise variance $v_b=7$.}
  \label{fig:jde_inputs}
\end{figure}

The generative model for ASL time series in section~\ref{sec:physio_JDE} has been used to generate artificial ASL data. A low SNR has been considered, with $TR = 1$~s, mean $ISI=5.03$~s, duration $25$~s, $N = 325$ scans and two experimental conditions ($M=2$) represented with $20  \times 20$-voxel binary activation label maps corresponding to BRL and PRL maps shown in Fig.~\ref{fig:jde_inputs}. For both conditions: ${(a_j^m|q_j=1) \sim {\cal N} (2.2, 0.3)}$ and ${(c_j^m|q_j=1) \sim {\cal N} (0.48, 0.1)}$. Parameters were chosen to simulate a typical low SNR ASL scenario, in which the perfusion component is much lower than the hemodynamics component.
A drift ${\ellb_j \sim {\cal N} (0,10\Ib_4)}$ and noise variance ${v_b = 7}$ were considered. 
BRF and PRF shapes were simulated with the physiological model, using the physiological parameters used in \cite{Friston00b}.

\begin{figure}[tbm]
\begin{center}
\btabu{c@{\hspace{1mm}}c@{\hspace{1mm}}c@{\hspace{0mm}}cc@{\hspace{0mm}}ccc}
(a) & & & (b) & & (c) & (d) & \vspace{0.1cm} \\
\figc[height=1.8cm]{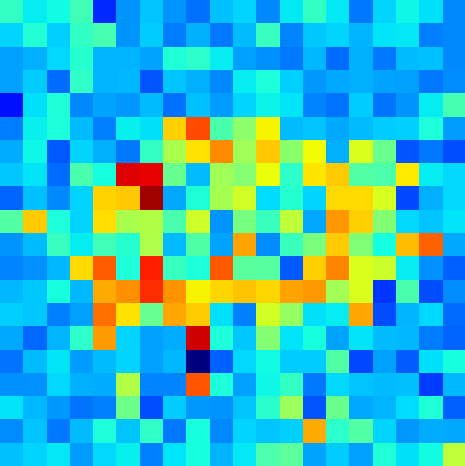} 
&\figc[height=1.8cm]{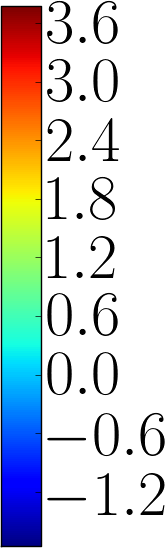}
&\yaxis{\tiny $\Delta$ BOLD signal}
& \figc[width=.25\linewidth]{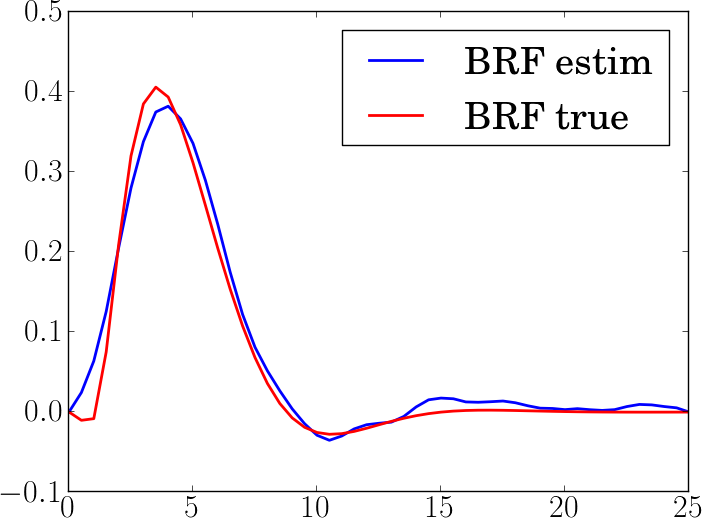} 
&\yaxis{\tiny $\Delta$ perfusion signal}
& \figc[width=.25\linewidth]{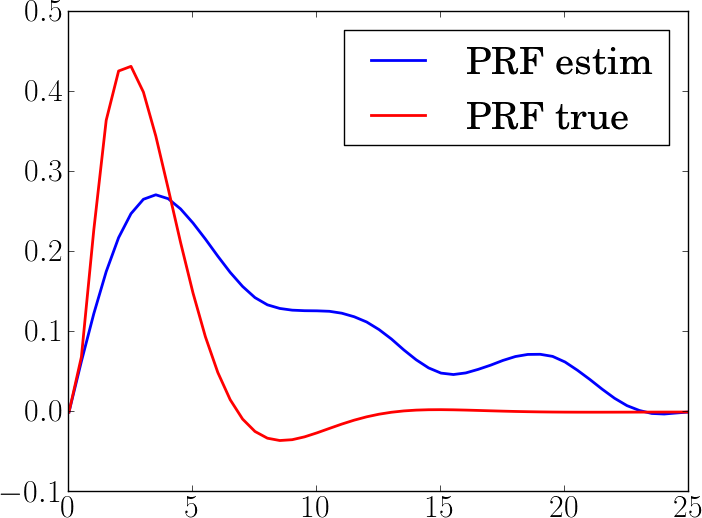} 
& \figc[height=1.8cm]{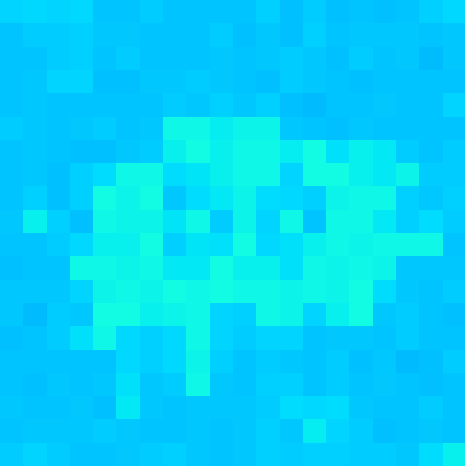}
&\figc[height=1.8cm]{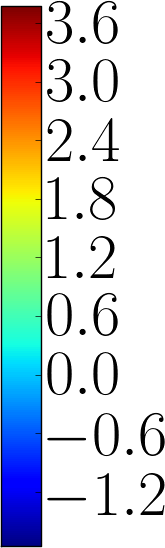}\\[-.15cm]
& & & {\tiny time (sec.)} & & {\tiny time (sec.)} & & \\[.2cm]
\figc[height=1.8cm]{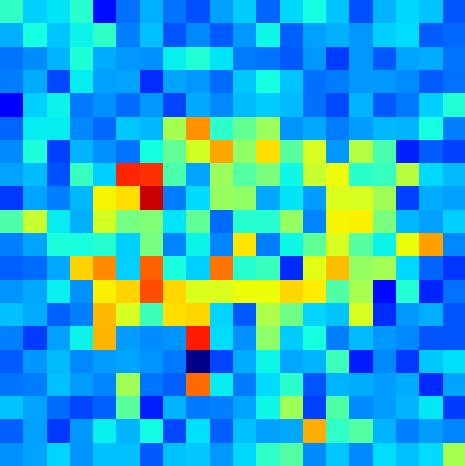} 
&\figc[height=1.8cm]{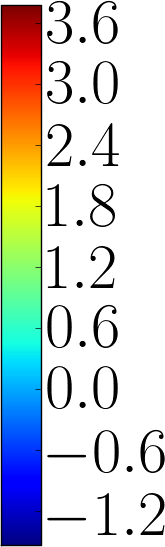}
&\yaxis{\tiny $\Delta$ BOLD signal}
& \figc[width=.25\linewidth]{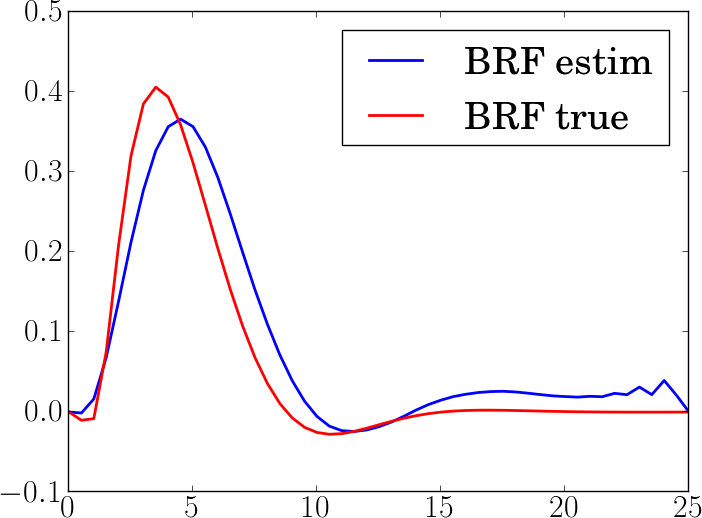} 
&\yaxis{\tiny $\Delta$ perfusion signal}
& \figc[width=.25\linewidth]{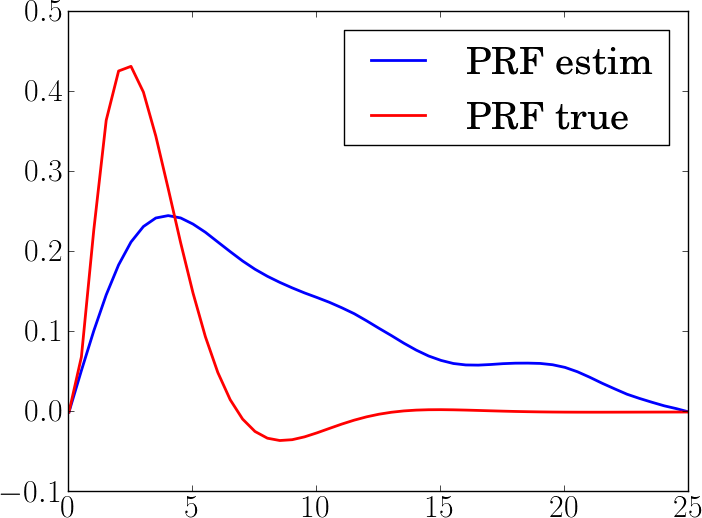} 
& \figc[height=1.8cm]{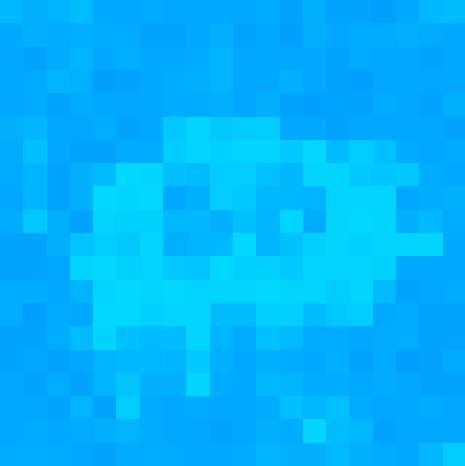}
&\figc[height=1.8cm]{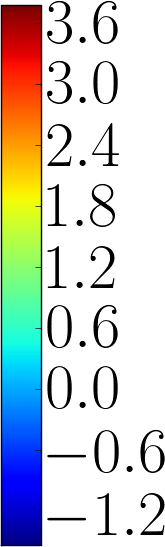}\\[-.15cm]
& & & {\tiny time (sec.)} & & {\tiny time (sec.)} & &
\\[.2cm]
\figc[height=1.8cm]{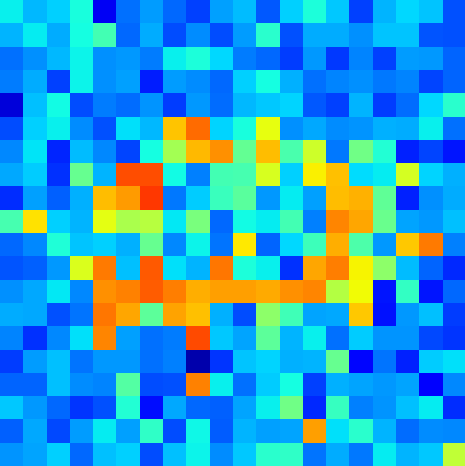} 
&\figc[height=1.8cm]{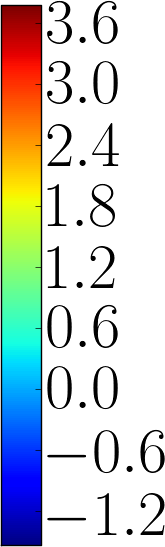}
&\yaxis{\tiny $\Delta$ BOLD signal}
& \figc[width=.25\linewidth]{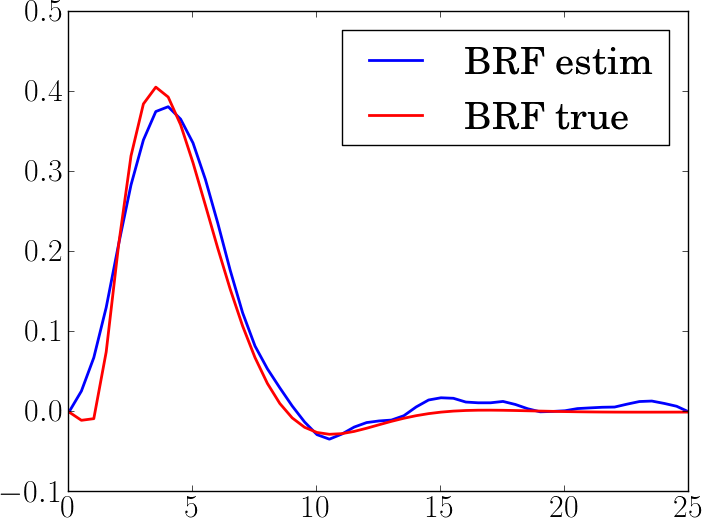} 
&\yaxis{\tiny $\Delta$ perfusion signal}
& \figc[width=.25\linewidth]{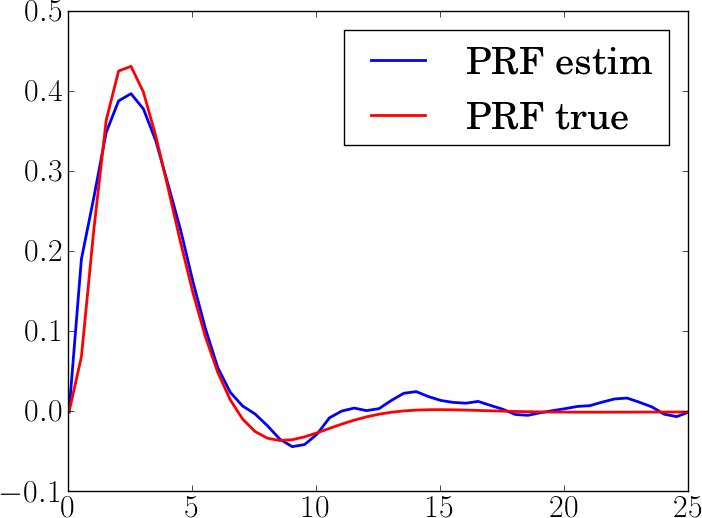} 
& \figc[height=1.8cm]{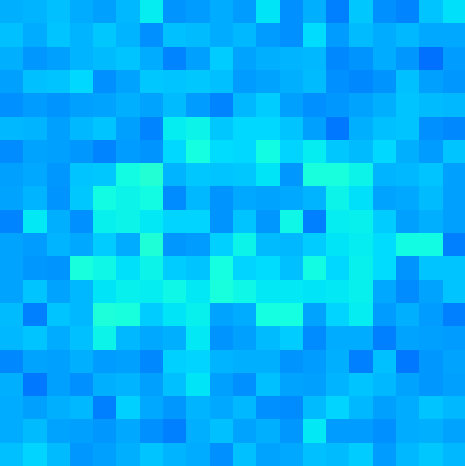}
&\figc[height=1.8cm]{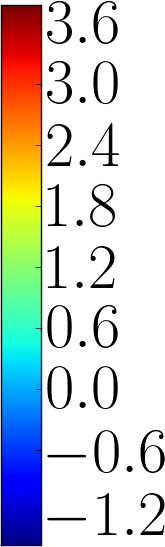}\\[-.15cm]
& & & {\tiny time (sec.)} & & {\tiny time (sec.)} & &
\etabu
\end{center}
\vspace{-.5cm}
\caption{Results on artificial data. \textbf{Top row}: non-physiological version. \textbf{Middle row}: physiological 1-step version. \textbf{Bottom row}: physiological 2-steps version. \textbf{(a,d)}: estimated BRL and PRL effect size maps respectively. The ground-truth maps for the BRL and PRL are depicted in~Fig.\ref{fig:jde_inputs}. \textbf{(b,c)}: BRF and PRF estimates, respectively, with their ground truth. 
\label{fig:jde_simulated}}
\end{figure}

In a low SNR context, the PRF estimate retrieved by the former approach developed in \cite{Vincent13b,Vincent13} is not physilogically relevant as shown in Fig.~\ref{fig:jde_simulated}[(c), Top]. In the case of a physiologically informed Bayesian approach, considering a single-step solution as in Fig.~\ref{fig:jde_simulated}[Middle], the perfusion component estimation is worse than for the approach described in \cite{Vincent13b,Vincent13} and  the BRF estimation is also degraded owing to the influence of the noisier perfusion component during the sampling.
In contrast, the 2-steps method proposed here delivers a PRF estimate very close to the simulated ground truth (see Fig.~\ref{fig:jde_simulated}[(c), Bottom] with a BRF which is well estimated too. 

\begin{figure}[tbm]
  \center
  \figc[width=0.5\textwidth]{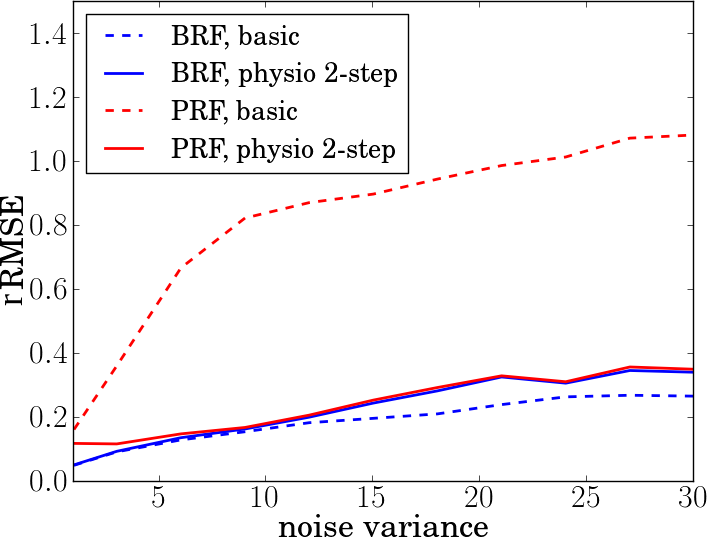}
  \caption{Relative RMSE for the BRF and PRF and the two JDE versions, wrt noise variance $v_b$ ranging from 0.5 to 30.}
  \label{fig:jde_simulated_noise}
\end{figure}

In Fig.~\ref{fig:jde_simulated_noise}, the robustness of both approaches with respect to the noise variance is studied, in terms of BRF and PRF recovery. The relative root-mean-square-error~(rRMSE) is computed for the PRF and BRF estimates, i.e. $\textrm{rRMSE}_{\phib} = \| \wh{\phib}- \phib^{(true)} \|/\|\phib^{(true)}\|$ where $\phib \in \{ \hb, \gb \}$. 
We observed that maintaining a good performance in the BRF estimation,
we achieved a much better recovery of the PRF  for noise variances larger than $v_b = 1$. Therefore, with the introduction of the physiological link between BRF and PRF, we have improved the PRF estimation.

\section{Real data results}
\label{sec:results_real}

\todo{R3- Validation is limited to one single ROI in a single subject only. Clearly, the PRF by design is better shaped according to what would be expected. Results appear to be scaled, without it being clear if this is an improvement. }

Real ASL data were recorded during an experiment designed
to map auditory and visual brain functions, which consisted of
 $N = 291$ scans lasting $TR = 3$~s, with $TE = 18$~ms, FoV $192$~mm, each
yielding a 3-D volume composed of $64 \times 64 \times 22$ voxels
(resolution of $3 \times 3 \times 3.5 $ mm$^3$). The tagging scheme used was PICORE Q2T, with $TI_1 = 700$~ms, $TI_2 = 1700$~ms.
The paradigm was a fast event-related design (mean $ISI=5.1$~s) comprising sixty auditory
and visual stimuli. Two regions of interest in the right temporal lobe, for the auditory cortex, and left occipital lobe, for the visual cortex, were defined manually.

\begin{figure}[tmb]
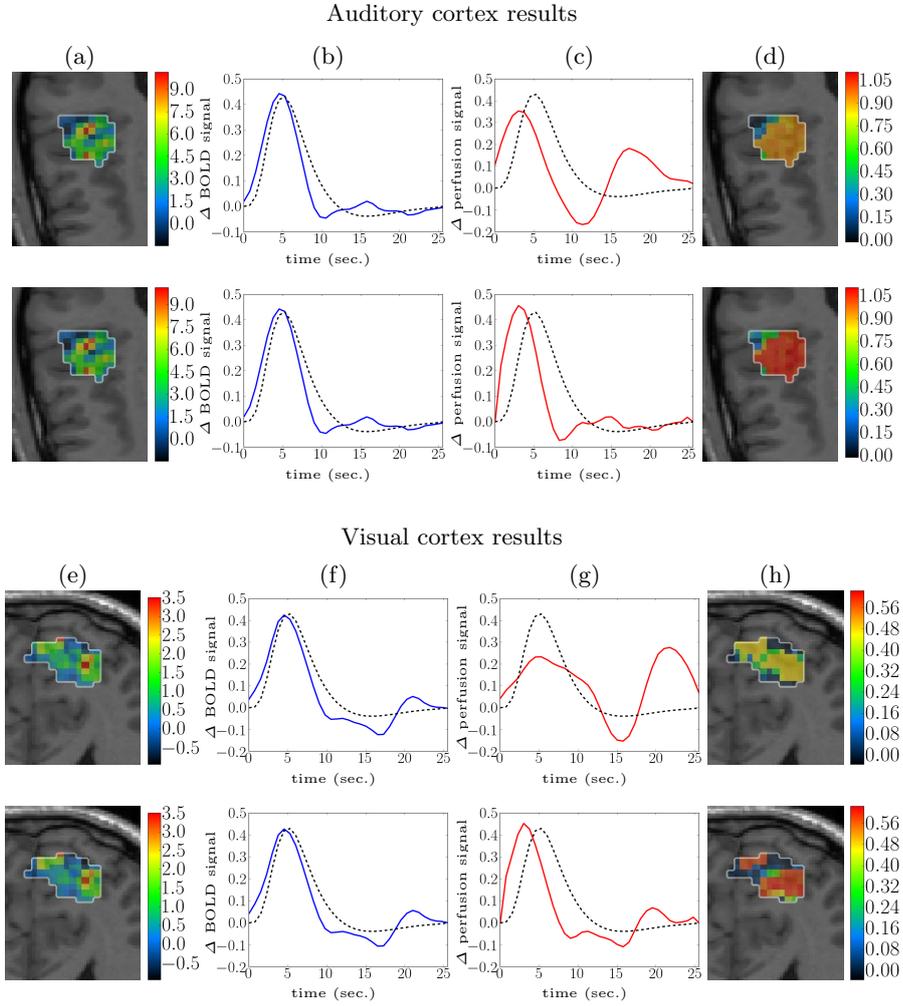

\begin{center}
Auditory cortex results
\\[.2cm]
\btabu{c@{\hspace{1mm}}c@{\hspace{1mm}}c@{\hspace{0mm}}cc@{\hspace{0mm}}ccc}
(a) & & & (b) & & (c) & (d) & \\
\figc[height=2.3cm]{audio_only/basic_real_data_jde_mcmc_brl_pm_cropped.png} 
&\figc[height=2.3cm]{audio_only/real_data_brls_palette.png}
&\yaxis{\tiny $\Delta$ BOLD signal}
& \figc[width=.25\linewidth]{audio_only/basic_real_data_brf.png} 
&\yaxis{\tiny $\Delta$ perfusion signal}
& \figc[width=.25\linewidth]{audio_only/basic_real_data_prf.png} 
& \figc[height=2.3cm]{audio_only/basic_real_data_jde_mcmc_prl_pm_cropped.png}
&\figc[height=2.3cm]{audio_only/real_data_prls_palette.png}\\[-.15cm]
& & & {\tiny time (sec.)} & & {\tiny time (sec.)} & & \\[.2cm]
\figc[height=2.3cm]{audio_only/2steps_real_data_jde_mcmc_brl_pm_cropped.png} 
&\figc[height=2.3cm]{audio_only/real_data_brls_palette.png}
&\yaxis{\tiny $\Delta$ BOLD signal}
& \figc[width=.25\linewidth]{audio_only/2steps_real_data_brf.png} 
&\yaxis{\tiny $\Delta$ perfusion signal}
& \figc[width=.25\linewidth]{audio_only/2steps_real_data_prf.png} 
& \figc[height=2.3cm]{audio_only/2steps_real_data_jde_mcmc_prl_pm_cropped.png} 
& \figc[height=2.3cm]{audio_only/real_data_prls_palette.png}\\[-.15cm]
& & & {\tiny time (sec.)} & & {\tiny time (sec.)} & & \\[.2cm]
\etabu
	\vspace{0.3cm}
Visual cortex results
\\[.2cm]
\btabu{c@{\hspace{1mm}}c@{\hspace{1mm}}c@{\hspace{0mm}}cc@{\hspace{0mm}}ccc}
(e) & & & (f) & & (g) & (h) & \\
\figc[height=2.3cm]{video_only/basic_real_data_jde_mcmc_brl_pm_cropped.png} 
&\figc[height=2.3cm]{video_only/real_data_brls_palette.png}
&\yaxis{\tiny $\Delta$ BOLD signal}
& \figc[width=.25\linewidth]{video_only/basic_real_data_brf.png} 
&\yaxis{\tiny $\Delta$ perfusion signal}
& \figc[width=.25\linewidth]{video_only/basic_real_data_prf.png} 
& \figc[height=2.3cm]{video_only/basic_real_data_jde_mcmc_prl_pm_cropped.png}
&\figc[height=2.3cm]{video_only/real_data_prls_palette.png}\\[-.15cm]
& & & {\tiny time (sec.)} & & {\tiny time (sec.)} & & \\[.2cm]
\figc[height=2.3cm]{video_only/2steps_real_data_jde_mcmc_brl_pm_cropped.png} 
&\figc[height=2.3cm]{video_only/real_data_brls_palette.png}
&\yaxis{\tiny $\Delta$ BOLD signal}
& \figc[width=.25\linewidth]{video_only/2steps_real_data_brf.png} 
&\yaxis{\tiny $\Delta$ perfusion signal}
& \figc[width=.25\linewidth]{video_only/2steps_real_data_prf.png} 
& \figc[height=2.3cm]{video_only/2steps_real_data_jde_mcmc_prl_pm_cropped.png} 
& \figc[height=2.3cm]{video_only/real_data_prls_palette.png}\\[-.15cm]
& & & {\tiny time (sec.)} & & {\tiny time (sec.)} & & \\[.2cm]
\etabu
\end{center}
\caption{Comparison of the two JDE versions on real data in the auditory and visual cortex. \textbf{(top row in auditory and visual cortex results)}: non-physiological version. \textbf{(bottom row in auditory and visual cortex results)}: physiological 2-steps version.  \textbf{(a,e)} and \textbf{(d,h)}: estimated BRL and PRL effect size maps, respectively. \textbf{(b,f)} and \textbf{(c,g)}: BRF and PRF estimates, respectively. The canonical BRF is depicted as a black dashed line, while PRF and BRF estimated are depicted in solid red and blue lines, respectively.
\label{fig:rfs_real_data}}
\end{figure}

Fig.~\ref{fig:rfs_real_data}(b-c) depicts the response estimates
superimposed to the canonical shape which is in accordance with the BRF estimates for both methods. Indeed, we consider here an auditory region where the canonical version has been fitted. Accordingly, the BRL maps (Fig.~\ref{fig:rfs_real_data}(a)) also look alike for both methods. However, PRF estimates significantly differ and the effect of the physiologically-inspired regularization yields a more plausible PRF shape for the 2-steps approach compared with the non-physiological JDE version. Results on PRL maps (Fig.~\ref{fig:rfs_real_data}(d)) confirm the improved sensitivity of detection for the proposed approach.
In the same way, in the visual cortex, Fig.~\ref{fig:rfs_real_data}(f-g) shows the BRF and PRF estimates, giving a more plausible PRF shape for the 2-steps approach, too. For the detection results (Fig.~\ref{fig:rfs_real_data}(h)), the 2-steps approach seems also to provide a much better sensitivity of detection.

\section{Discussion and conclusion}
\label{sec:discussion}

Starting from non-linear systems of differential equations induced by physiological models of the neuro-vascular coupling, we derived a tractable linear operator linking the perfusion and  BOLD responses. This operator showed good approximation performance and demonstrated its ability to capture both realistic perfusion and BOLD components. 
In addition, this derived linear operator was easily incorporated in a JDE framework at no additional cost and with a  significant improvement in PRF estimation, especially in critical low SNR situations.  As shown on simulated data, the PRF estimation has been improved while maintaining accurate BRF estimation. Real data results seem to confirm the better performance of the proposed physiological approach
compared to its competing alternative. In terms of validation, future work will be devoted to intensive validation on whole brain analysis and multiple subjects. 



\bibliographystyle{splncs}
\bibliography{myBib}

\begin{thebibliography}{10}

\bibitem{Williams92}
Williams, D., Detre, J., Leigh, J., Koretsky, A.:
\newblock Magnetic resonance imaging of perfusion using spin inversion of
  arterial water.
\newblock Proceedings of the National Academy of Sciences \textbf{89}(1) (1992)
   212--216

\bibitem{Ogawa92}
Ogawa, S., Tank, D., Menon, R., Ellermann, J., Kim, S.G., Merkle, H., Ugurbil,
  K.:
\newblock Intrinsic signal changes accompanying sensory stimulation: functional
  brain mapping with magnetic resonance imaging.
\newblock {P}roc. {N}atl. {A}cad. {S}ci. {USA} \textbf{89} (1992)  5951--5955

\bibitem{Hernandez10}
Hernandez-Garcia, L., Jahanian, H., Rowe, D.B.:
\newblock Quantitative analysis of arterial spin labeling fmri data using a
  general linear model.
\newblock Magnetic resonance imaging \textbf{28}(7) (2010)  919--927

\bibitem{Mumford06}
Mumford, J.A., Hernandez-Garcia, L., Lee, G.R., Nichols, T.E.:
\newblock Estimation efficiency and statistical power in arterial spin labeling
  fmri.
\newblock Neuroimage \textbf{33}(1) (2006)  103--114

\bibitem{Vincent09c}
Vincent, T., Risser, L., Ciuciu, P.:
\newblock Spatially adaptive mixture modeling for analysis of {within-subject
  fMRI} time series.
\newblock {IEEE} {T}rans. {M}ed. {I}mag. \textbf{29}(4) (April 2010)
  1059--1074

\bibitem{Vincent13b}
Vincent, T., Warnking, J., Villien, M., Krainik, A., Ciuciu, P., Forbes, F.:
\newblock {Bayesian Joint Detection-Estimation of cerebral vasoreactivity from
  ASL fMRI data}.
\newblock In: 16th {P}roc. MICCAI, LNCS Springer Verlag. Volume~2., Nagoya,
  Japan (September 2013)  616--623

\bibitem{Vincent13}
Vincent, T., Forbes, F., Ciuciu, P.:
\newblock Bayesian {BOLD} and perfusion source separation and deconvolution
  from functional {ASL} imaging.
\newblock In: 38th {P}roc. {IEEE} {I}CASSP, Vancouver, Canada (May 2013)
  1003--1007

\bibitem{Friston00b}
Friston, K.J., Mechelli, A., Turner, R., Price, C.J.:
\newblock Nonlinear responses in f{MRI}: the balloon model, {V}olterra kernels,
  and other hemodynamics.
\newblock {N}euroimage \textbf{12} (June 2000)  466--477

\bibitem{Buxton04}
Buxton, R.B., Uluda{\u{g}}, K., Dubowitz, D.J., Liu, T.T.:
\newblock Modeling the hemodynamic response to brain activation.
\newblock Neuroimage \textbf{23} (2004)  S220--S233

\bibitem{Buxton98}
Buxton, R.B., Wong, E.C., R., F.L.:
\newblock Dynamics of blood flow and oxygenation changes during brain
  activation: the balloon model.
\newblock {M}agn. {R}eson. {M}ed. \textbf{39} (June 1998)  855--864

\bibitem{Khalidov11}
Khalidov, I., Fadili, J., Lazeyras, F., Van De~Ville, D., Unser, M.:
\newblock Activelets: {W}avelets for sparse representation of hemodynamic
  responses.
\newblock Signal Processing \textbf{91}(12) (December 2011)  2810--2821

\bibitem{Makni06b}
Makni, S., Ciuciu, P., Idier, J., Poline, J.B.:
\newblock Bayesian joint detection-estimation of brain activity using {MCMC
  }with a {Gamma-Gaussian} mixture prior model.
\newblock In: 31th {P}roc. {IEEE} {I}CASSP. Volume~V., Toulouse, France (May
  2006)  1093--1096

\end{thebibliography}

\end{document}